\documentclass[12pt,smallextended,natbib,runningheads,epsfig]{article}
\usepackage{latexsym, amsfonts}
\usepackage{amssymb, amsmath}
\usepackage{amsfonts}
\usepackage{amsfonts}
\usepackage{graphicx}
\usepackage{adjustbox}
\usepackage{setspace}
\usepackage{rotating}
\usepackage{tikz}
\usepackage{bigints}
\usepackage{bm}
\usepackage{multirow}
\usepackage{subcaption}

\newcommand{\be}{\begin{equation}}
\newcommand{\ee}{\end{equation}}
\newcommand{\bea}{\begin{eqnarray}}
\newcommand{\eea}{\end{eqnarray}}
\newcommand{\bean}{\begin{eqnarray*}}
\newcommand{\eean}{\end{eqnarray*}}
\newcommand{\brray}{\begin{array}}
\newcommand{\erray}{\end{array}}
\newcommand{\ben}{\begin{equation}{nonumber}}
\newcommand{\een}{\end{equation}{nonumber}}

\newtheorem{dfn}{Definition}[section]
\newtheorem{thm}[dfn]{Theorem}
\newtheorem{lmma}[dfn]{Lemma}
\newtheorem{ppsn}[dfn]{Proposition}
\newtheorem{crlre}[dfn]{Corollary}
\newtheorem{xmpl}[dfn]{Example}
\newtheorem{rmrk}[dfn]{Remark}

\newcommand{\bdfn}{\begin{dfn}}
\newcommand{\bthm}{\begin{thm}}
\newcommand{\blmma}{\begin{lmma}}
\newcommand{\bppsn}{\begin{ppsn}}
\newcommand{\bcrlre}{\begin{crlre}}
\newcommand{\bxmpl}{\begin{xmpl}}
\newcommand{\brmrk}{\begin{rmrk}}
\newcommand{\edfn}{\end{dfn}}
\newcommand{\ethm}{\end{thm}}
\newcommand{\elmma}{\end{lmma}}
\newcommand{\eppsn}{\end{ppsn}}
\newcommand{\ecrlre}{\end{crlre}}
\newcommand{\exmpl}{\end{xmpl}}
\newcommand{\ermrk}{\end{rmrk}}

\def\a*{{\cal A}_{h,*}}
\def\B{{\cal B}(h)}
\def\B1{{\cal B}_1(h)}
\def\b{{\cal B}^{\rm s.a.}(h)}
\def\b1{{\cal B}^{\rm s.a.}_1(h)}

\numberwithin{equation}{section}

\begin{document}
\begin{center}
{\Large {\bf A Latent Class Bayesian Model for Multivariate Longitudinal Outcomes with Excess Zeros}}\\

\noindent Chitradipa Chakraborty$^{1}$ and Kiranmoy Das$^{2}$  \\
\
\noindent{$^1$ Beijing Key Laboratory of Topological Statistics and Applications for Complex Systems, Beijing Institute of Mathematical Sciences and Applications, Beijing, China}\\
\noindent{$^2$ Statistical Science Division, Indian Statistical Institute, Kolkata, India}
\end{center}

\begin{abstract}
Latent class models have been successfully used to handle complex datasets in different disciplines. For longitudinal outcomes, we often get a trajectory of the outcome for each individual, and on that basis, we cluster them for a powerful statistical inference. Latent class models have been used to handle multivariate longitudinal outcomes coming from biology, health sciences, and economics. In this paper, we propose a Bayesian latent class model for multivariate outcomes with excess zeros. We consider a Tobit model for zero-inflated continuous outcomes such as out-of-pocket medical expenses (OOPME), a two-part model for financial debt, and a ZIP model for counting outcomes with excess zeros. We develop a Bayesian mixture model and employ an adaptive Lasso-type shrinkage method for variable selection. We analyze data from the Health and Retirement Study conducted by the University of Michigan and consider modeling four important outcomes measuring the physical and financial health of the aged individuals. Our analysis detects several latent clusters for different outcomes. Practical usefulness of the proposed model is validated through a simulation study. 
\vspace{0.2 in}

\end{abstract}

\vspace{0.1 in}

\section{Introduction}
Clustering is an essential technique for gaining a better understanding of the data at hand and enabling efficient statistical inference. For longitudinal outcomes, where we get a trajectory for each subject, clustering is done based on the individual trajectories. Typically, there are two approaches for clustering longitudinal outcomes, i.e. (i) model-based approach and (ii) algorithm-based approach. 
In a model-based approach one assumes a finite mixture model and each of the mixing components corresponds to one cluster. The model-based approach includes group-based trajectory models proposed in Nagin (1999), Nagin \& Odgers (2010). This approach assumes that individuals belonging to a cluster have the same trajectory. As an alternative, latent class mixed effect model, proposed in Muthén \& Shedden (1999), Proust-Lima et al. (2017), assumes that the subject-specific trajectories vary around a common mean trajectory. Model-based approach assumes a probability distribution for the data and the inference heavily depends on the assumed probability distribution. The algorithm-based approach, on the other hand, assumes no probability distribution for the dataset and uses computational algorithms for clustering the data. The methods include $k$-means clustering (Genolini et al. (2016)), hierarchical clustering (Zhou et al. (2023)) and clustering based on correlation (Pinto da Costa et al. (2021)).  

While both the model-based and the algorithm-based methods are quite popular and effective for clustering univariate longitudinal outcomes, computation becomes challenging for clustering multivariate longitudinal outcomes. In addition, multivariate longitudinal outcomes might be complex due to missing observations, zero-inflation and time-varying dependence among the outcomes. In this paper, we address the problem of zero inflation, where some of the longitudinal outcomes indeed contain excess zeros. Data containing excess zeros are not uncommon in medical, biological or economic datasets. There is a rich literature on zero-inflated longitudinal data (Biswas and Das (2021), Das, Ghosh and Daniels (2021)) and the references therein. 

Our work in this paper is motivated by the Health and Retirement Study (HRS), conducted by the University of Michigan, to understand the challenges of aging. The study started in 1992 with a longitudinal cohort born between 1931 and 1941, and ended in 2012. Data were collected from the subjects on their physical health, financial health and demographics every two years referred to as waves. The HRS dataset has been analyzed by Brown et al. (2015) who proposed a two-part model to study the financial debt of the aged individuals. Bhuyan et al. (2018) developed a two-stage regression model for simultaneously handling the endogeneity and missingness in the data. Biswas and Das (2020) proposed a Bayesian approach for predicting the out-of-pocket medical expenditures (OOPME) for the aged individuals. Similar models were proposed in Mukherji et al. (2016). 
HRS dataset contains four variables which summarize the physical health and financial health of the aged individuals. These variables are Out-of Pocket Medical Expenditures (OOPME), total financial assets, total financial debts and the number of hospital stays in a specific wave. We note that except the total financial assets, all the other variables contain excess zeros which is not  counter-intuitive. The dataset, in addition, contains a large number of covariates and hence variable selection is a natural component of our analysis. We use the Adaptive Lasso (Zou (2006)) within a Bayesian framework, following the methodology proposed by Leng et al. (2010). 

To account for the heterogeneity in the dataset a latent class model is used to explore subgroups, cluster longitudinal features, and investigate cluster-cluster similarities and dissimilarities in the outcomes, based on a Bayesian Mixture Model (BMM). We perform cluster-specific variable selection and the computation is performed using Markov Chain Monte Carlo (MCMC) algorithm in RJAGS. Our analysis finds two latent clusters with different patterns in the trajectories of the longitudinal outcomes. 

The remainder of the paper is organized as follows. Section 2 details the proposed modeling framework, the application of a latent class model for clustering longitudinal features, the handling of zero-inflated responses, and encompassing the Bayesian Adaptive Lasso for variable selection. Section 3 presents the empirical analysis of the HRS dataset, and interprets the findings. Section 4 summarizes the findings from a simulation study. Some concluding remarks are given in Section 5. 

\section{Model and Method}

Let $y_{irj}$ denote the longitudinal response for the $i$th individual, $r$th feature, and $j$th observation, where $i = 1, \ldots, m$, $r = 1, \ldots, R$, and $j = 1, \ldots, n_{ir}$. For the entire study population, a Generalized Linear Mixed Model (GLMM) for each response can be written as
\[
\boldsymbol{y}_{ir} = \boldsymbol{X}_{ir}\boldsymbol{\beta}_r + \boldsymbol{z}_{ir}\boldsymbol{\gamma}_{ir} + \boldsymbol{\epsilon}_{ir},
\]
where $\boldsymbol{y}_{ir} = (y_{ir1}, \ldots, y_{irn_{ir}})^\top$, $\boldsymbol{X}_{ir}$ is the covariate matrix, and $\boldsymbol{z}_{ir}$ is the design vector for random effects. The covariate matrix is defined as $\mathbf{X}_{ir} = \left( x^{(1)}_{ir}, \ldots, x^{(P)}_{ir} \right)$, where each 
$x^{(p)}_{ir} = \left( x^{(p)}_{ir1}, \ldots, x^{(p)}_{irn_{ir}} \right)^\top \in \mathbb{R}^{n_{ir}}$ 
is a column vector corresponding to the $p$-th covariate, for $p = 1, \ldots, P$. Generally, $y_{irj}$ can be any type of response, such as continuous, discrete, or categorical, recorded at a given time point. The fixed effects $\boldsymbol{\beta}_r$ represent a vector of unknown regression coefficients, while the random effects $\boldsymbol{\gamma}_{ir}$ are subject- and feature-specific coefficients that account for within-subject correlation. The residuals $\boldsymbol{\epsilon}_{ir}$ are assumed to be independently distributed as $N(0, \sigma^2 \mathbf{I})$.

\subsection{Bayesian Mixture Model} 
For clustering multiple longitudinal features, the underlying assumption is that the study population is heterogeneous and composed of K latent classes of individuals characterized by distinct trajectory patterns of multiple features. The Bayesian Mixture Model (BMM) is built upon a multivariate mixture of generalized linear mixed models (Kom\'arek and Kom\'arkov\'a, 2013). In this framework, the mixture distribution is applied to the random effects only, not the entire response distribution. Let $\boldsymbol{\gamma}_i=(\boldsymbol{\gamma}_{i1}^T,\ldots, \boldsymbol{\gamma}_{iR}^T)^T$ denote the joint vector of random effects for individual $i$, where the total dimension is $q=\displaystyle\sum_{r=1}^Rq_r$. This structure allows the model to capture the correlation among the $R$ longitudinal features. Instead of modeling the entire outcome vector $\boldsymbol{y}_i=(\boldsymbol{y}_{i1},\ldots,\boldsymbol{y}_{iR})^T$ directly, clustering is induced through a finite mixture model on the random effects: $$P(\boldsymbol{\gamma}_i|\boldsymbol{\theta}) = \displaystyle\sum_{k=1}^K \pi_k P(\boldsymbol{\gamma}_i|\boldsymbol{\theta}_k),$$ where $\pi_k \in [0,1]$ are mixing proportions satisfying $\displaystyle\sum_{k=1}^K \pi_k = 1.$ If $C_i \in \{1,\ldots,K\}$ denotes the latent class (cluster) membership for individual $i$, then $\pi_k=P(C_i=k),$ and the cluster assignment follows a multinomial distribution: $C_i \sim \text{Multinomial} (1;\boldsymbol{\pi}), \text{ with }  \boldsymbol{\pi}=(\pi_1,\ldots,\pi_K).$ In this model, $\boldsymbol{\theta} =(\boldsymbol{\pi}, \boldsymbol{\theta}_1,\ldots,\boldsymbol{\theta}_K)$ and each component distribution $P(\boldsymbol{\gamma}_i|\boldsymbol{\theta}_k)$ is assumed to be a multivariate normal distribution with parameters  $\boldsymbol{\theta}_k = (\boldsymbol{\mu}_k, \boldsymbol{\Psi}_k)$, where $\boldsymbol{\mu}_k$ is a $q$ dimension mean vector and $\boldsymbol{\Psi}_k$ is a $q \times q$ variance-covariance matrix corresponding to cluster $k$. For model identifiability, the covariates $\boldsymbol{X}_{ir}$ and $\boldsymbol{z}_{ir}$ do not contain the same variables. The model is estimated using a MCMC algorithm, and individuals are assigned to the cluster for which their posterior membership probability is highest. The optimal number of clusters ($K$) is selected based on the lowest penalized expected deviance (PED), calculated by incorporating the expected deviance of the model, with an additional penalty for model complexity (optimism).

\subsection{Zero-Inflated Responses}
In many applications, response variables are nonnegative and exhibit a positive probability of being exactly zero. When a variable follows a continuous distribution but has a point mass at zero, it is referred to as semi-continuous. Semi-continuous data are common across various domains, and unlike left-censored data, the zero values reflect true observed outcomes. Similarly, zero-inflated count data arise when there are more zeros than would be expected under standard models such as the Poisson. Analyzing these types of data poses challenges, as the presence of excess zeros renders common distributions like the normal, gamma, or standard count models (e.g., Poisson) inadequate, often resulting in poor model fit. These characteristics motivate the need for specialized modeling approaches.

Tobin (1958) proposed a censored regression model, now widely known as the Tobit model, to describe household expenditures on durable goods. Cowles, Carlin, and Connett (1996) extended the classical Tobit model to longitudinal data. The model is specified as:

\[
y_{irj} = \begin{cases}
y_{irj}^*, & \text{if } y_{irj}^* > 0 \\
0, & \text{if } y_{irj}^* \leq 0
\end{cases}
\]

\noindent The latent variable is modeled using:
$$y_{irj}^* = \boldsymbol{x}_{irj}'\boldsymbol{\beta}_r + \boldsymbol{z}_{irj}'\boldsymbol{a}_{ir} + \epsilon_{irj},$$
where the residuals $\epsilon_{irj} \sim N(0, \sigma^2)$. 

The Tobit model assumes normally distributed errors with constant variance, an assumption that may be unrealistic in many practical settings. 
To address this, Duan et al. (1983) proposed a two-part model that separates the modeling process into two stages: the first stage models the probability of a nonzero response, and the second stage models the outcomes conditional on being positive. Olsen and Schafer (2001) further extended this framework to accommodate longitudinal data. In the first part of the model, the probability of a zero or nonzero outcome is modeled using a logistic mixed-effects framework. Specifically, 
\[
y_{irj}  \begin{cases}
= 0, & \text{with probability } p_{irj} \\
\neq 0, & \text{with probability  } 1-p_{irj}
\end{cases}
\]

\noindent Let $\{\boldsymbol{a}_{ir}\}$ be random effects to account for within-subject correlation. Conditional on $\boldsymbol{a}_{ir}$, it is assumed that 
$$\text{logit}(p_{irj}) = \boldsymbol{x}_{1irj}'\boldsymbol{\beta} _{1r} + \boldsymbol{z}_{1irj}'\boldsymbol{a}_{ir}.$$ 

\noindent In the second part of the model, let
\[
V_{irj} =  \begin{cases}
 y_{irj}, & \text{if } y_{irj} > 0 \\
\text{unspecified,} & \text{if  } y_{irj} = 0
\end{cases}
\]
When $V_{irj}$ is positive, conditional on a random effect $\boldsymbol{b}_{ir}$ the model
assumes that $V_{irj}$ follows a log-normal distribution. Thus, the model
for the positive outcomes is
$$\text{log}(V_{irj}) = \boldsymbol{x}_{2irj}'\boldsymbol{\beta}_{2r} + \boldsymbol{z}_{2irj}'\boldsymbol{b}_{ir} + \epsilon_{irj},$$ where $\epsilon_{irj} \sim N(0, \sigma^2)$.

For zero-inflated count data, Lambert (1992) introduced zero-inflated Poisson (ZIP) regression models to account for overdispersion in the form of excess zero counts for the Poisson distribution. Lambert’s model treats the data as a mixture of zeros and outcomes of Poisson variates. Hall (2000) extended the ZIP model to accommodate longitudinal data by incorporating random effects to capture within-subject correlation. In Hall’s formulation, the response variable follows 
\[
Y_{irj}  \begin{cases}
= 0, & \text{with probability } p_{irj} \\
\sim \text{Poisson}(\lambda_{irj}), & \text{with probability  } 1-p_{irj}
\end{cases}
\] 
with the parameters modeled by $$\text{logit}(p_{irj}) = \boldsymbol{x}_{1irj}'\boldsymbol{\beta} _{1r} \quad \text{and} \quad \text{log}(\lambda_{irj}) = \boldsymbol{x}_{2irj}'\boldsymbol{\beta} _{2r} + \boldsymbol{z}_{2irj}'\boldsymbol{d}_{ir},$$ where $\boldsymbol{d}_{ir}$ is a random effect capturing individual-level variation. Yau and Lee (2001) further extended this framework by introducing random effects into both components of the model, that is, $$\text{logit}(p_{irj}) = \boldsymbol{x}_{1irj}'\boldsymbol{\beta} _{1r} + \boldsymbol{z}_{2irj}'\boldsymbol{c}_{ir} \quad \text{and} \quad \text{log}(\lambda_{irj}) = \boldsymbol{x}_{2irj}'\boldsymbol{\beta} _{2r} + \boldsymbol{z}_{2irj}'\boldsymbol{d}_{ir},$$ where $\boldsymbol{c}_{ir}$ and $\boldsymbol{d}_{ir}$ represent random effects for the zero-inflation and count components, respectively.

\subsection{Bayesian Adaptive Lasso for Variable Selection}

Tibshirani (1996) introduced the Least Absolute Shrinkage and Selection Operator (Lasso), which performs simultaneous variable selection and parameter estimation by imposing an $l_1$-penalty on the regression coefficients. In a Bayesian framework, the Lasso estimator can be viewed as the posterior mode under a normal likelihood with independent Laplace priors on the regression coefficients $\beta_j$. However, Fan and Li (2001) demonstrated that although the Lasso is capable of automatic variable selection, it tends to produce biased estimates for large coefficients, and thus does not possess the oracle property. To address this limitation, Zou (2006) proposed the Adaptive Lasso, which introduces data-driven weights into the penalty function and satisfies the oracle property under certain regularity conditions. 

In the Bayesian context, Park and Casella (2008) developed the Bayesian Lasso (BLasso), which enables full posterior inference by treating the penalized regression as a hierarchical model. This approach was further extended by Kyung et. al. (2010), who proposed a more general Bayesian hierarchical framework capable of encompassing a range of penalized regression methods, including the group Lasso, the fused Lasso, and the elastic net. Extending these ideas further, Leng et al. (2010) introduced the Bayesian Adaptive Lasso (BaLasso), which provides a unified approach for variable selection with flexible penalty structures. Their method is based on the least squares approximation strategy (Wang and Leng, 2007), and allows the penalty to adapt to individual coefficients. By assigning different levels of shrinkage to different coefficients through a hierarchical structure, BaLasso offers greater flexibility in variable selection and estimation. Specifically, the Bayesian Adaptive Lasso assumes the following prior hierarchy:
\[
\beta_r^{(j)} \mid \lambda_j \sim \text{Laplace}(0, \lambda_j^{-1})
\]
\[
\text{or equivalently,} \quad
\beta_r^{(j)} \mid \tau_j^2 \sim \mathcal{N}(0, \tau_j^2), \quad
\tau_j^2 \sim \text{Exp}(\lambda_j^2 / 2)
\]
\[
\text{with} \quad \lambda_j^2 \sim \text{Gamma}(a, b), \quad \text{for} \quad j=1,\ldots,P,
\]
where $P$ denotes the total number of covariates. This hierarchical formulation allows for adaptive, coefficient-specific shrinkage, thereby improving model flexibility and enabling more accurate variable selection.

\section{Data Analysis}

We analyze data from 1,418 individuals, each observed across 10 time points, resulting in a balanced longitudinal dataset with $n_{ir} = n = 10$ for all individuals $i$ and response variables $r$. Our focus is on four longitudinal responses ($R = 4$): out-of-pocket medical expenditures (OOPME), total financial assets, total financial debt, and the number of hospital stays. Among these, OOPME, financial assets, and financial debt are continuous variables, while the number of hospital stays is a count variable. OOPME exhibits 8.3\% zero observations, and total financial debt has 39\% zeros, necessitating specialized modeling approaches. We model OOPME using an extended Tobit model and total financial debt using a two-part model that separately addresses the probability of a nonzero response and the conditional distribution given positive values. The number of hospital stays, with 86\% of observations being zero, is modeled using an extended zero-inflated Poisson (ZIP) model to accommodate both overdispersion and excess zeros. In the model framework, the wave indicator is included as a random-effect covariate vector $\boldsymbol{z}_{ir}$, while the fixed-effects design matrix $\boldsymbol{X}_{ir}$ comprises the remaining 20 covariates. We apply a Bayesian Adaptive Lasso prior on the regression coefficients to perform variable selection among these covariates. \\

\begin{figure}[htbp]
  \centering

  \begin{minipage}[b]{0.32\linewidth}
    \centering
    \includegraphics[width=\linewidth]{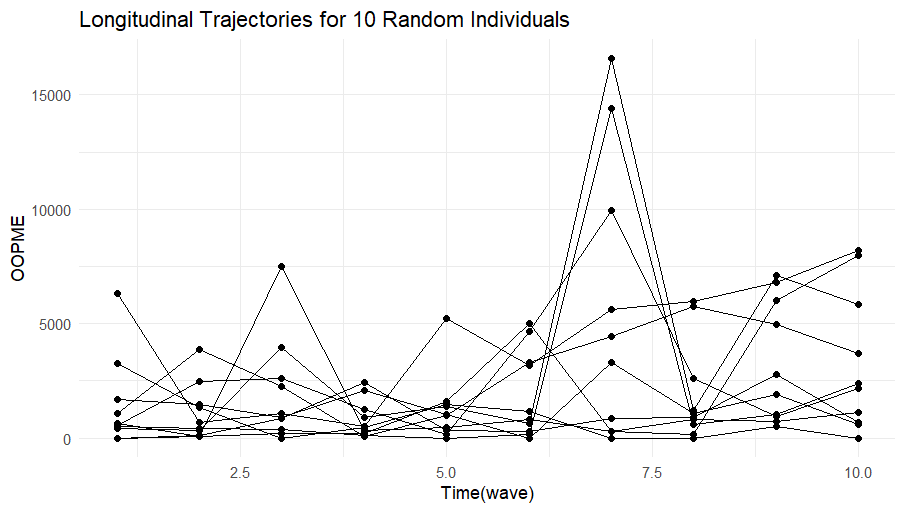}
    \subcaption{}
    \label{fig:sub1}
  \end{minipage}
  \hfill
  \begin{minipage}[b]{0.32\linewidth}
    \centering
    \includegraphics[width=\linewidth]{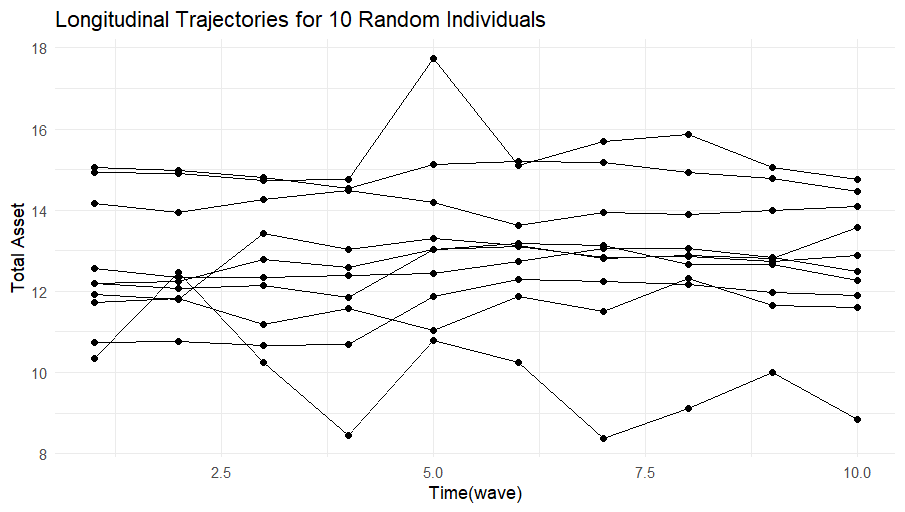}
    \subcaption{}
    \label{fig:sub2}
  \end{minipage}
  \hfill
  \begin{minipage}[b]{0.32\linewidth}
    \centering
    \includegraphics[width=\linewidth]{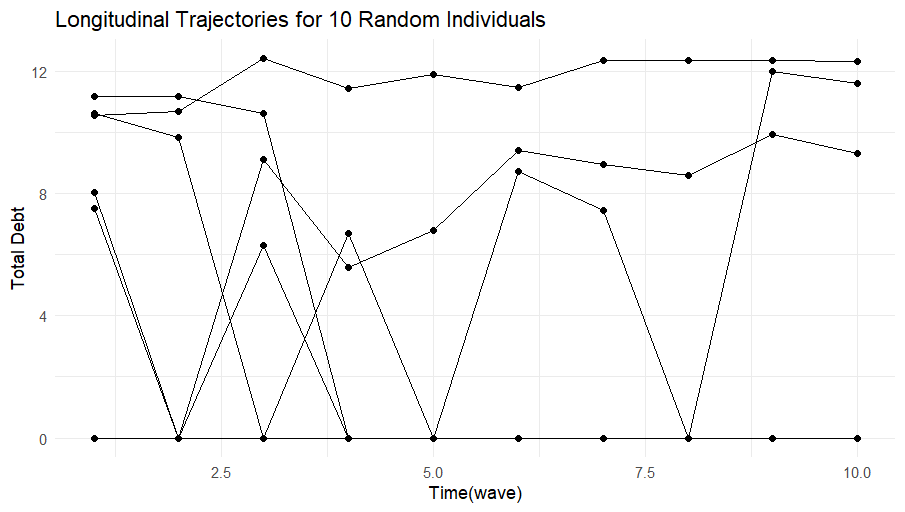}
    \subcaption{}
    \label{fig:sub3}
  \end{minipage}

  \caption{Mean longitudinal trajectories of three features for 10 randomly selected individuals in HRS dataset. (a) longitudinal trajectories of OOPME, (b) longitudinal trajectories of Total Financial Assets, (c) longitudinal trajectories of Total Financial Debt.}
  \label{fig:trajectories}
\end{figure}

The plots in Figure 1 illustrate the distinct patterns observed in the longitudinal data, showing how individuals exhibit varying trajectories in out-of-pocket medical expenditures (OOPME), total financial assets, and total financial debt over time. These differing patterns suggest that the population is heterogeneous, with subgroups (latent clusters) potentially experiencing different economic and health trajectories. The variability in the data implies that a single model might not adequately capture the complexity of the underlying processes, highlighting the need for latent clusters to identify and separate these distinct subgroups. By clustering individuals based on their similar longitudinal trajectories, we can better understand the patterns and underlying dynamics of financial and health outcomes across the population. A latent class model, based on the Bayesian Mixture Model (BMM), is employed to explore these subgroups. To determine the optimal number of clusters ($K$), we computed the PED for a range of $K$ values. The optimal $K$ was selected as the value of K that minimized the PED, as this indicated the best model with respect to both fit and complexity. \\

\begin{table}[h!]
\centering
\begin{tabular}{|c|c|}
\hline
\textbf{$\boldsymbol{K}$} & \textbf{PED} \\
\hline
2 & 442623.9\\
\hline
3 & 443090.7 \\
\hline
4 & 452140.6 \\
\hline
5 & 447782.0 \\
\hline
\end{tabular}
\caption{The PED values for different numbers of clusters.}
\end{table}

As shown in Table 1, the optimal number of clusters, $K = 2$, corresponds to the minimum Penalized Expected Deviance (PED = 442623.9). This indicates that a two-cluster solution best balances model fit and complexity. Following this selection, the variance-covariance matrices for the random effects, denoted as \(\boldsymbol{\Psi}_k\) (\(k = 1, 2\)), are estimated. These matrices are of size \(6 \times 6\), corresponding to the dimensionality of the random effects, and are essential for understanding the correlation structure between the random effects, which is important for identifying how various health and financial variables interact within each cluster.. The estimated values of these matrices are presented below:\\

\begin{figure}[htbp]
\centering
\[
\boldsymbol{\Psi}_1 =
\begin{bmatrix}
24.905 & -7.350 & -2.153 &  2.106 &  3.392 & -2.600 \\
-7.350 & 88.820 &  0.070 & -0.312 &  0.549 & -1.217 \\
-2.153 &  0.070 & 14.800 & -0.391 & -0.093 & -0.429 \\
 2.106 & -0.312 & -0.391 & 61.345 &  1.532 & -1.830 \\
 3.392 &  0.549 & -0.093 &  1.532 & 16.999 &  0.094 \\
-2.600 & -1.217 & -0.429 & -1.830 &  0.094 & 15.774
\end{bmatrix}
\]
\end{figure}

\begin{center}
    and
\end{center}

\begin{figure}[htbp]
\centering
\[
\boldsymbol{\Psi}_2 =
\begin{bmatrix}
10.982 & -2.549 &  0.347 & -0.050 & -0.430 &  0.152 \\
-2.549 & 12.565 &  0.094 & -0.712 & -0.768 &  0.618 \\
 0.347 &  0.094 &  9.110 & -0.132 &  0.663 &  0.366 \\
-0.050 & -0.712 & -0.132 &  9.289 &  0.429 &  0.669 \\
-0.430 & -0.768 &  0.663 &  0.429 &  9.942 & -0.579 \\
 0.152 &  0.618 &  0.366 &  0.669 & -0.579 &  8.444
\end{bmatrix}
\]

\label{fig:phi_matrices}
\end{figure}


Figure 2 represents the cluster-specific longitudinal trajectories that reveal distinct economic patterns. In analyzing out-of-pocket medical expenditures (OOPME), two distinct clusters emerge. Cluster 1 (Stable) shows a gradual upward trend with moderate fluctuations from the early 1990s through the 2010s, indicating that most households experience a steady increase in medical expenses over time. Conversely, Cluster 2 (Volatile) reveals a sharp spike around wave 3, followed by large oscillations, suggesting that a small subset of households faces episodic, severe medical cost shocks. Regarding total financial assets, Cluster 1 (Stable) displays consistently high asset levels, with a slow but steady increase\\

\begin{figure}[htbp]
  \centering

  \begin{minipage}[b]{0.32\linewidth}
    \centering
    \includegraphics[width=\linewidth]{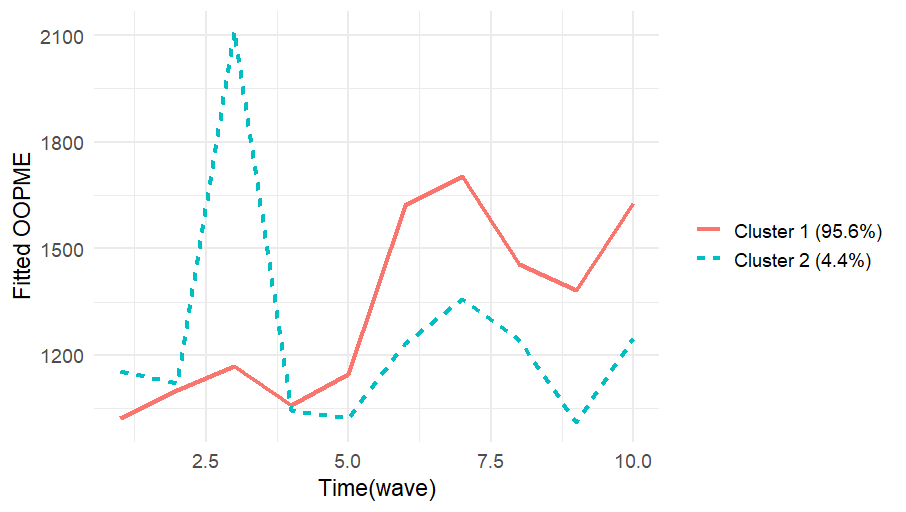}
    \subcaption{}
    \label{fig:sub1}
  \end{minipage}
  \hfill
  \begin{minipage}[b]{0.32\linewidth}
    \centering
    \includegraphics[width=\linewidth]{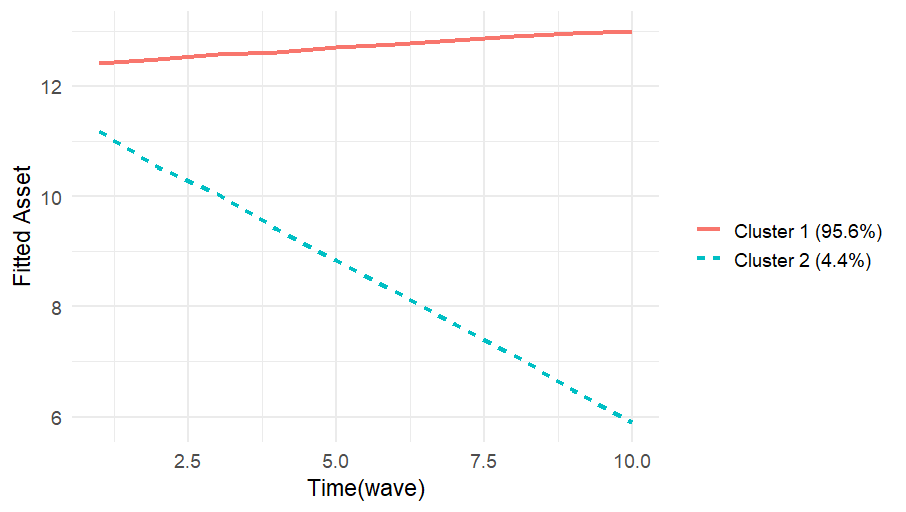}
    \subcaption{}
    \label{fig:sub2}
  \end{minipage}
  \hfill
  \begin{minipage}[b]{0.32\linewidth}
    \centering
    \includegraphics[width=\linewidth]{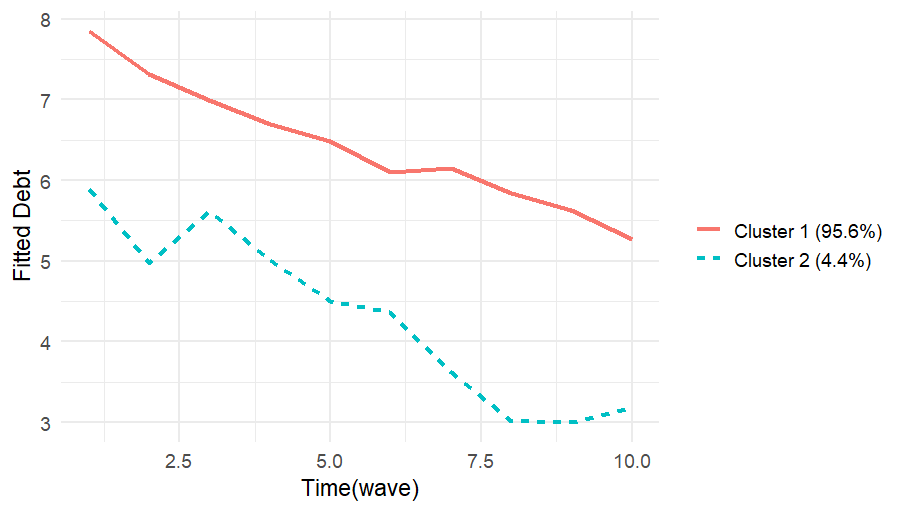}
    \subcaption{}
    \label{fig:sub3}
  \end{minipage}

  \caption{Longitudinal trajectories of three features by clusters. (a) longitudinal trajectories of OOPME, (b) longitudinal trajectories of Total Financial Assets, (c) longitudinal trajectories of Total Financial Debt.}
  \label{fig:trajectories}
\end{figure}

\noindent across waves, indicating that wealthier households are able to accumulate assets over time. In contrast, Cluster 2 (Declining) starts with moderately high assets but experiences a steady decline, reaching nearly half by wave 10, highlighting a vulnerable group facing asset depletion. Finally, in terms of total financial debt, Cluster 1 (Stable) shows a higher baseline debt, but with a gradual decline across waves, reflecting that the majority of households are steadily reducing their debt. Cluster 2 (Volatile), on the other hand, starts with lower overall debt, experiencing fluctuations early on but a sharp decline after wave 5, suggesting that a smaller group is reducing debt more rapidly through sharper contractions. Overall, these patterns reveal a clear divide between wealthier households, which accumulate assets and reduce debt steadily, and a more vulnerable minority facing significant financial instability, with fluctuating medical expenditures and declining assets.
 
\begin{table}[ht]
\centering
\begin{tabular}{lcccc}
\hline
  & \textbf{Hospital Stays} & \textbf{OOPME} & \textbf{Total Assets} & \textbf{Total Debt} \\
\hline
Age                         & \checkmark & \checkmark & \checkmark  &  \checkmark   \\
Gender                      & $\times$   & $\times$   & $\times$   & \checkmark   \\
Alcohol Consumption         & $\times$   & $\times$   & \checkmark   & $\times$   \\
Smoking Habit               &  \checkmark   & $\times$   & \checkmark  & $\times$   \\
Education Level             & $\times$   & $\times$   & \checkmark   & \checkmark   \\
Work Limitation             &  \checkmark  & $\times$   & \checkmark  & $\times$   \\
Hypertension                &  \checkmark   & $\times$   & \checkmark  & \checkmark   \\
Diabetes                   & $\times$   & $\times$   & $\times$   & \checkmark   \\
Cancer                     & $\times$   & $\times$   & \checkmark   & $\times$   \\
Lung Problem               &  \checkmark   & $\times$   & $\times$   & $\times$   \\
Heart Problem              & $\times$   & $\times$   & \checkmark   & $\times$   \\
Stroke                     & $\times$   & $\times$   & $\times$   & $\times$   \\
Arthritis                  & $\times$   & $\times$   & $\times$   & \checkmark \\
Psychological Problems     & $\times$   & $\times$   & $\times$   & \checkmark   \\
Self-Assessed Health Status & \checkmark & $\times$   & \checkmark    & \checkmark   \\
No Insurance               &  \checkmark   & $\times$   & \checkmark   & \checkmark  \\
Employment Insurance       &  \checkmark   & $\times$   & \checkmark   & \checkmark   \\
Government Insurance       &  \checkmark   & $\times$   & \checkmark   & \checkmark   \\
Other Insurance            &  \checkmark  & $\times$   & \checkmark  & \checkmark  \\
Household Income     & $\times$   & $\times$   & \checkmark   & \checkmark   \\
\hline
\end{tabular}
\caption{Predictor inclusion in outcome models: checkmark indicates predictor selected/important, cross indicates not selected.}
\label{tab:predictors}
\end{table}

From Table 2, we can observe that the hospital stays are primarily influenced by age, smoking, health conditions (e.g., hypertension, lung issues), work limitations, and insurance coverage—factors indicative of both access to and need for hospital care. OOPME is predominantly determined by age alone, possibly due to insurance coverage minimizing variation across individuals. Total assets are shaped by a broad set of factors including demographics, behaviors, health conditions, and insurance, underscoring the multifactorial nature of long-term wealth accumulation. In contrast, total debt is driven by socioeconomic variables such as gender, income, education, and insurance status, as well as some health conditions. Age consistently emerges as a key determinant across all outcomes, linking physical aging with both health-related and financial dimensions. Financial indicators such as assets and debt are more sensitive to socioeconomic variables, whereas health indicators more directly affect hospital utilization. Notably, OOPME exhibits less variability in predictors, likely reflecting systemic factors such as insurance coverage that buffers against high expenses. 

To understand the relationship between different outcomes (such as medical expenditures, assets, and debts) over time, we can estimate the expected value of one outcome conditioned on the values of other outcomes while leaving the covariates out of the equation. This process involves calculating the conditional mean of the outcome, which is done by integrating over the covariates using their empirical distributions. In other words, we account for the variability in the covariates based on their observed frequency or distribution in the data. By plotting the conditional means for each wave, we gain insights into how the outcome of interest (e.g., out-of-pocket medical expenditures, total financial assets, or debt levels) behaves over time, conditional on specific values of other outcomes, without the complicating influence of individual covariates. This approach helps us focus on the relationship between the key variables across different time periods and identify trends, peaks, or fluctuations that might be masked by covariate variations. Essentially, we’re isolating the dynamics between the variables of interest and observing their evolution, free from the noise introduced by factors like income, age, or employment status.\\

\begin{figure}[htbp]
  \centering
    \includegraphics[width=0.9\linewidth]{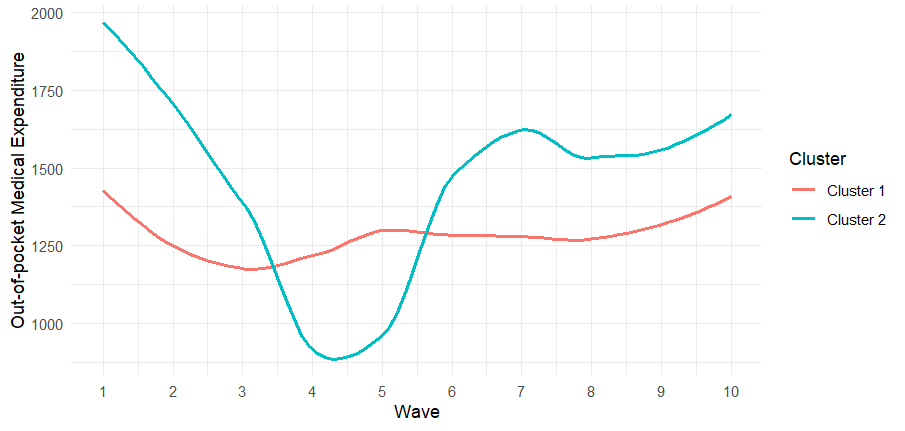}
    \caption{Mean out-of-pocket medical expenditures by latent clusters conditional on the third quartile and above of the total assets, first quartile and below of total debts, and the number of hospital admissions $\geq 3$, and with government insurance coverage.}
    \label{fig:oopme_clusters}
  \end{figure}
  
The analysis of medical expenditures across different household characteristics reveals contrasting patterns depending on insurance status, assets, debts, and hospitalization frequency. For households with high assets, low debts, and three or more hospitalizations (Figure 3), those in Cluster 1 show steady out-of-pocket medical expenditures (OOPME) from 1992 to 2012, indicating a predictable financial burden despite the presence of government insurance. However, Cluster 2 experiences very high OOPME in 1992, followed by a sharp drop in the mid-1990s, then a rebound through the 2000s and 2010s. This suggests that government insurance does not fully shield this group from medical cost fluctuations, \\

\begin{figure}[htbp]
  \centering
    \includegraphics[width=0.9\linewidth]{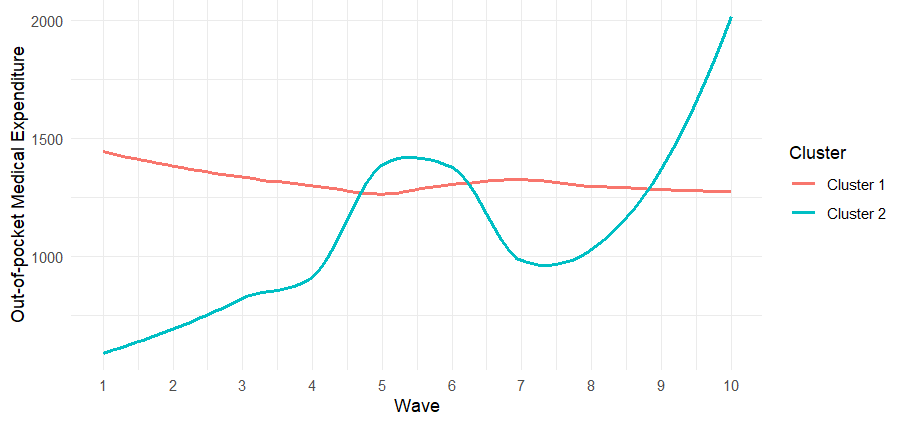}
    \caption{Mean out-of-pocket medical expenditures by latent clusters conditional on the first quartile and below of the total assets, first quartile and below of total debts, and the number of hospital admissions $\geq 3$, and with employment insurance coverage.}
    \label{fig:oopme_clusters}
  \end{figure}

\noindent particularly in periods of high need. For households with low assets, low debts, and three or more hospitalizations covered by employment insurance (Figure 4), Cluster 1 shows high but steady expenditures with a slight downward drift, indicating that employment insurance provides consistent protection, though not without a long-term upward trend in medical costs. Meanwhile, Cluster 2 experiences very low OOPME in the early 1990s, but sees steady increases by 2010–2012, revealing that while employment insurance initially offers relief, medical costs become progressively more burdensome over time. For households with high assets, low debts, very few hospitalizations, and no insurance (Figure 5), Cluster 1 presents moderate and stable OOPME, reflecting a predictable cost burden in the absence of insurance, while Cluster 2 sees fluctuations—dipping in the mid-1990s, rebounding in the 2000s, dipping again in the mid-2000s, and then stabilizing in the 2010s. This variability suggests that the lack of insurance leads to episodic financial stress for some households, with medical costs rising unpredictably in response to health events.

\begin{figure}[htbp]
  \centering
    \includegraphics[width=0.9\linewidth]{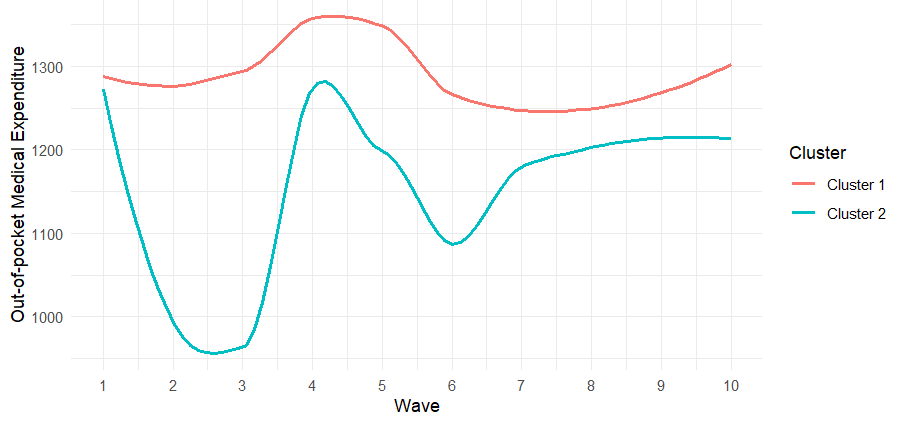}
    \caption{Mean out-of-pocket medical expenditures by latent clusters conditional on the third quartile and above of the total assets, first quartile and below of total debts, and the number of hospital admissions $\leq 2$, and with no insurance coverage.}
    \label{fig:oopme_clusters}
  \end{figure}

In analyzing the mean total financial assets across different household characteristics, we observe distinct patterns depending on insurance coverage and medical expenses. For households with low debts, high out-of-pocket medical expenditures (OOPME), and very few hospitalizations covered by government insurance (Figure 6), Cluster 1 shows a fairly flat trajectory with minor fluctuations, suggesting that while government insurance offers some stability, it doesn't significantly affect asset growth or depletion. In contrast, Cluster 2 experiences strong swings in asset levels, with notable peaks in the mid-2000s and a sharp dip around wave 8, followed by a rebound by wave 10, indicating that government insurance partially cushions financial shocks but cannot fully protect against the volatility of medical expenses. For households with high OOPME, low debts, and very few hospitalizations under employment insurance (Figure 7), Cluster 1 demonstrates a gradual decline in assets over time, reflecting the long-term erosion of financial stability despite initial protection\\

\begin{figure}[htbp]
    \centering
    \includegraphics[width=0.9\linewidth]{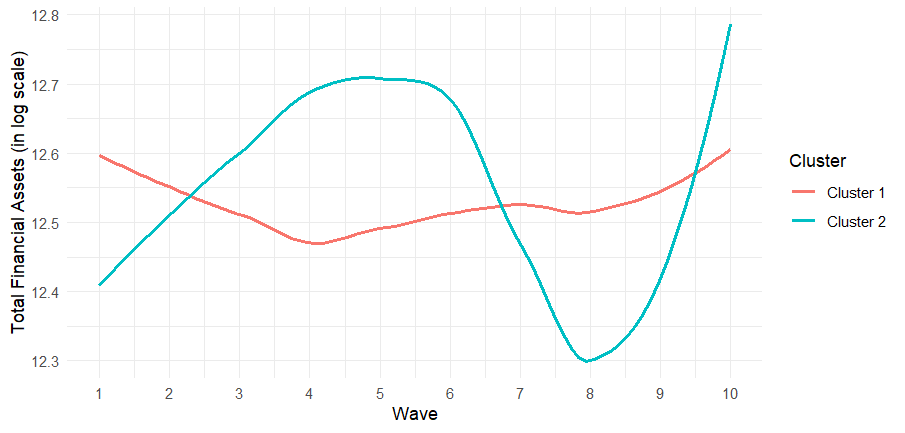}
    \caption{Mean total financial assets by latent clusters conditional on the first quartile and below of the total debt, the third quartile and above of the OOPME, and the number of hospital admissions $\leq 2$, and with government insurance coverage.}
    \label{fig:asset_clusters}
\end{figure}

\begin{figure}[htbp]
    \centering
    \includegraphics[width=0.9\linewidth]{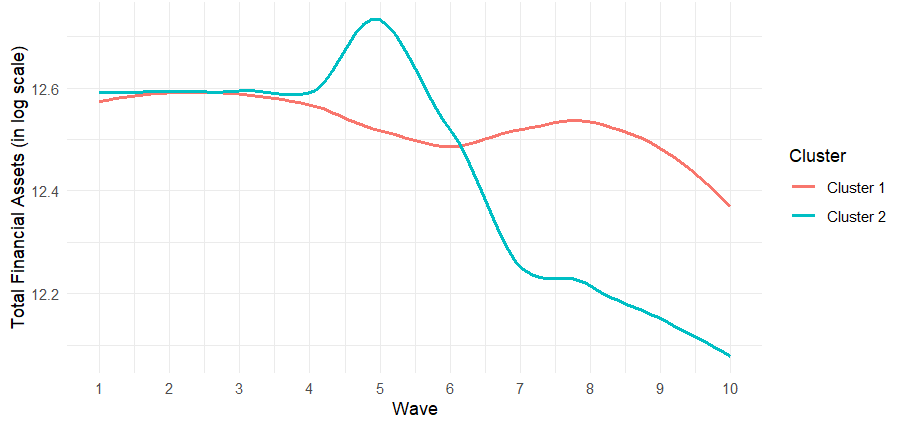}
    \caption{Mean total financial assets by latent clusters conditional on the third quartile and above of the OOPME, first quartile
and below of total debts, and the number of hospital admissions $\leq 2$, and with employment insurance coverage.}
    \label{fig:asset_clusters}
\end{figure}

\noindent from employment-based coverage. Cluster 2 sees a rise in assets up until wave 5, followed by a steady decline through wave 10, showing that while employment insurance may provide temporary relief, the escalating medical costs eventually outweigh the benefits. Finally, for households with high OOPME, high debts, and three or more hospitalizations but no insurance (Figure 8), Cluster 1 shows a predictable and flat asset trajectory, indicating that despite high debts and medical expenses, these households manage to maintain asset levels, possibly through other financial mechanisms. However, Cluster 2 experiences significant peaks between waves 4 and 7, followed by a sharp crash by wave 10, illustrating that the lack of insurance leads to severe financial instability and sharp fluctuations in asset levels. Overall, these patterns reveal that while insurance provides some degree of protection, it does not fully prevent asset depletion, especially for households facing high medical expenses or lacking adequate coverage.

\begin{figure}[htbp]
    \centering
    \includegraphics[width=\linewidth]{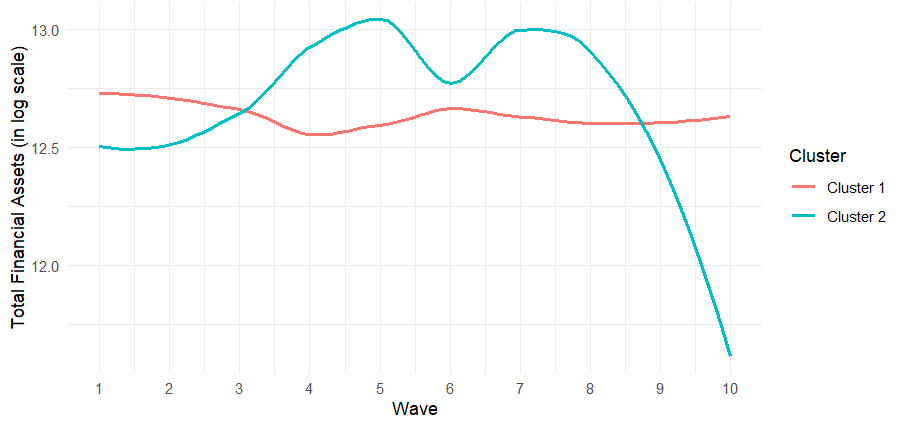}
    \caption{Mean total financial assets by latent clusters conditional on the third quartile and above of the OOPME, third quartile
and above of total debts, and the number of hospital admissions $\geq 3$, and with no insurance coverage.}
    \label{fig:asset_clusters}
\end{figure}


Figure 9 presents trace and posterior density plots for the precision parameters $\tau_{\text{oopme}}$, $\tau_{\text{lntotass}}$, and $\tau_{\text{lntotdebt}}$, each evaluated across three MCMC chains. The trace plots demonstrate good mixing, with the chains fluctuating around stable means and exhibiting no signs of non-convergence such as persistent trends or high autocorrelation. The density plots are smooth and unimodal, further supporting that the posterior distributions are well explored and stably estimated. The Gelman-Rubin convergence diagnostics ($\hat{R}$) for these parameters are 1.011 for $\tau_{\text{oopme}}$, 1.002 for $\tau_{\text{lntotass}}$, and 1.003 for $\tau_{\text{lntotdebt}}$, all well below the standard convergence threshold of 1.1 (Gelman and Rubin, 1992). These results indicate that the MCMC chains have converged adequately and that the posterior estimates for the precision of out-of-pocket medical expenses, log total assets, and log total debt are robust and reliable.\\

\begin{figure}[htbp]
  \centering
  \includegraphics[width=\linewidth]{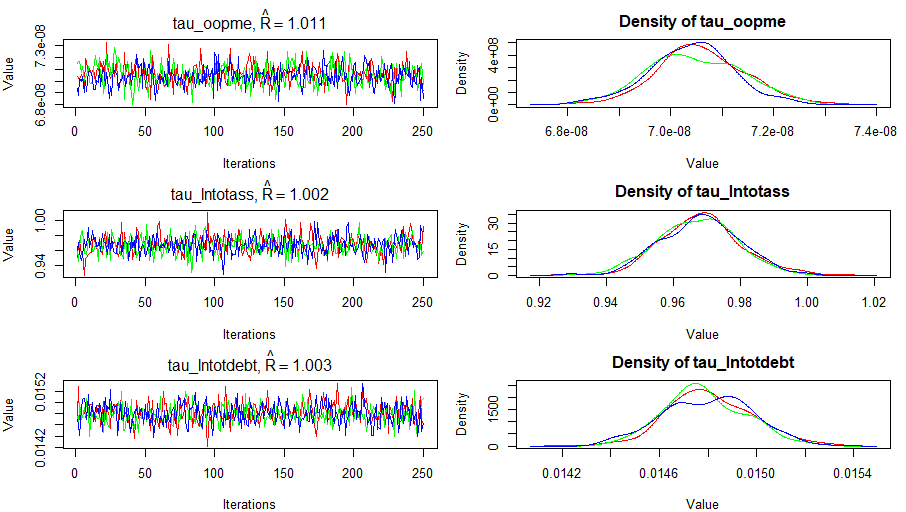}
  \caption{Trace and Density plots for $\tau_{\text{oopme}}$, $\tau_{\text{lntotass}}$, and $\tau_{\text{lntotdebt}}$}
  \label{fig:tau_trace}
\end{figure}

\section{Simulation Study}

In our simulation study, we consider four covariates, among which two covariates with values substantially higher or lower than zero, and two covariates with values close to zero. We simulate the longitudinal outcomes, with one outcome being zero-inflated for 300 individuals, each observed at 10 evenly spaced time points (\( t = 1, 2, \dots, 10 \)). Each individual is assigned to one of two latent clusters (\( K = 2 \)), with 70\% belonging to cluster 1 and 30\% to cluster 2. The random effects \( R = (R_1, R_2) \) are generated according to a Bayesian Mixture Model (BMM) with the following mixture of normal distributions:

\[
R \sim 0.7 N(\boldsymbol{0}, I) + 0.3 N(\boldsymbol{1}, 1.5I).
\]

\noindent The resulting random effects matrix \( R \) is of size \( 3000 \times 2 \), corresponding to the two longitudinal responses for each individual.

The first longitudinal outcome, \( y_1 \), is generated using a linear mixed-effects model:

\[
y_1 = X^{T}\beta + Z R_1 + \epsilon, 
\]

\noindent and the second longitudinal outcome, \( y_2 \), is zero-inflated, with 30\% of the values being zero. We use a two-part model for this response, defined as:

\[
y_2 = \begin{cases} 
0 & \text{with probability 0.3} \\
y^* & \text{with probability 0.7, where} \quad y^* = X^{T}\beta + ZR_2 + \epsilon
\end{cases}
\]

\noindent In both cases, \( \beta \) is the fixed effect coefficient vector, \( Z \) is the time covariate, which accounts for the repeated measurements over time, and random error $\epsilon \sim N(0, 1)$.\\
\begin{table}[h!]
\centering
\begin{tabular}{|c|c|}
\hline
\textbf{$\boldsymbol{K}$} & \textbf{PED} \\
\hline
2 & 55504.00 \\
\hline
3 & 55509.18\\
\hline
4 & 55526.88 \\
\hline
5 & 55530.25 \\
\hline
\end{tabular}
\caption{The PED values for different numbers of clusters.}
\end{table}

Using the simulated data, we apply the Bayesian Mixture Model (BMM) to assess the ability of the model to recover the latent clusters. 
As shown in Table 3, the optimal number of clusters, K = 2, corresponds to the minimum Penalized Expected Deviance (PED = 55504). This indicates that model accurately identifies the two clusters.

\section{Concluding Remarks}
Zero-inflation brings additional complexity to multivariate longitudinal datasets, particularly when dealing with a large number of covariates whose effects vary across different clusters. In this work, we developed a computationally efficient Bayesian model for clustering such datasets and applied it to the Health and Retirement Study (HRS) dataset. The analysis revealed two distinct latent clusters with markedly different economic and health trajectories. One cluster displayed relatively stable patterns, with steadily increasing medical expenses, consistently high financial assets, and gradually declining debt, suggesting greater financial security and resilience. The other cluster, in contrast, exhibited volatile medical expenditures, declining assets, and unstable debt trajectories, highlighting a more vulnerable group prone to financial and health shocks. These findings underscore the usefulness of the proposed model in uncovering meaningful latent classes within complex longitudinal data. The effectiveness of the model was further supported through a simulation study, where it successfully identified the correct cluster structure and accounted for zero-inflated responses. In addition, the Bayesian adaptive Lasso framework allowed for covariate selection, emphasizing the importance of factors such as age, smoking habits, and insurance coverage in shaping both health and financial outcomes.

While this work provides an important step toward modeling complex zero-inflated longitudinal data, extensions remain possible. In particular, missing data, which is common in longitudinal studies, was not addressed here. For ignorable missingness (i.e., missing at random), a data-augmentation step could be incorporated within the MCMC iterations to simultaneously update model parameters and missing observations, while non-ignorable missingness can be handled using approaches such as those suggested by Daniels and Hogan (2008). These aspects will form the focus of future work.


\begin{thebibliography}{99}
\setlength{\itemsep}{0pt}
\setlength{\parskip}{0pt}

\bibitem{Bhuyan2018} 
Bhuyan, P., Biswas, J., Ghosh, P., \& Das, K. (2018). A Bayesian two-stage regression approach of analysing longitudinal outcomes with endogeneity and incompleteness. \textit{Statistical Modelling}, \textbf{19}(2), 157--173.

\bibitem{Biswas2020} 
Biswas, J., \& Das, K. (2020). A Bayesian approach of analyzing semi-continuous longitudinal data with monotone missingness. \textit{Statistical Modelling}, \textbf{20}, 148--170.

\bibitem{Biswas2021} 
Biswas, J., \& Das, K. (2021). A Bayesian quantile regression approach to multivariate semi-continuous longitudinal data. \textit{Computational Statistics}, \textbf{36}, 241--260.

\bibitem{Brown2015} 
Brown, S., Ghosh, P., Su, L., \& Taylor, K. (2015). Modelling household finances: A Bayesian approach to a multivariate two-part model. \textit{Journal of Empirical Finance}, \textbf{33}, 190--207.

\bibitem{Cowles1996} 
Cowles, M. K., Carlin, B. P., \& Connett, J. E. (1996). Bayesian Tobit modeling of longitudinal ordinal clinical trial compliance data with nonignorable missingness. \textit{Journal of the American Statistical Association}, \textbf{91}, 86--98.

\bibitem{Daniels2008} 
Daniels, M. J., \& Hogan, J. W. (2008). \textit{Missing data in longitudinal studies: Strategies for Bayesian modeling and sensitivity analysis} (1st ed.). Chapman and Hall/CRC. 

\bibitem{Das2021} 
Das, K., Ghosh, P., \& Daniels, M. J. (2021). Modeling multiple time-varying related groups: A dynamic hierarchical Bayesian approach with an application to the Health and Retirement Study. \textit{Journal of the American Statistical Association}, \textbf{116}(534), 558--568.

\bibitem{Duan1983} 
Duan, N., Manning, W. G. Jr., Morris, C. N., \& Newhouse, J. P. (1983). A comparison of alternative models for the demand for medical care (Corr: V2 P413). \textit{Journal of Business and Economic Statistics}, \textbf{1}, 115--126.

\bibitem{Fan2001} 
Fan, J., \& Li, R. (2001). Variable selection via nonconcave penalized likelihood and its oracle properties. \textit{Journal of the American Statistical Association}, \textbf{96}(456), 1348--1360.

\bibitem{gelman1992}
Gelman, A. \& Rubin, D. B. (1992). Inference from Iterative Simulation
Using Multiple Sequences. \textit{Statistical Science}, \textbf{7}(4), 457-511.

\bibitem{Genolini2016} 
Genolini, C., Ecochard, R., Benghezal, M., Driss, T., Andrieu, S., \& Subtil, F. (2016). kmlShape: an efficient method to cluster longitudinal data (time-series) according to their shapes. \textit{PLOS One}, \textbf{11}(6), e0150738.

\bibitem{Hall2000} 
Hall, D. B. (2000). Zero-inflated Poisson and binomial regression with random effects: a case study. \textit{Biometrics}, \textbf{56}, 1030--1039.

\bibitem{Komarek2013} 
Komárek, A., \& Komárková, L. (2013). Clustering for multivariate continuous and discrete longitudinal data. \textit{Annals of Applied Statistics}, \textbf{7}(1), 177--200.

\bibitem{Kyung2009} 
Kyung, M., Gill, J., Ghosh, M., \& Casella, G. (2010). Penalized regression, standard errors and Bayesian Lassos. \textit{Bayesian Analysis}, \textbf{5}(2), 369--412.

\bibitem{Lambert1992} 
Lambert, D. (1992). Zero-inflated Poisson regression, with an application to defects in manufacturing. \textit{Technometrics}, \textbf{34}, 1--14.

\bibitem{Leng2014} 
Leng, C., Tran, M. N., \& Nott, D. (2010). Bayesian adaptive lasso. \textit{Annals of the Institute of Statistical Mathematics}, \textbf{66}(2), 221--244.

\bibitem{Mukherji2016} 
Mukherji, A., Roychoudhury, S., Ghosh, P., \& Brown, S. (2016). Estimating health demand for an aging population: A flexible and robust Bayesian joint model. \textit{Journal of Applied Econometrics}, \textbf{31}, 1140--1158.

\bibitem{Muthén1999} 
Muth\'en, B., \& Shedden, K. (1999). Finite mixture modeling with mixture outcomes using the EM algorithm. \textit{Biometrics}, \textbf{55}(2), 463--469.

\bibitem{Nagin1999} 
Nagin, D. S. (1999). Analyzing developmental trajectories: a semiparametric, group-based approach. \textit{Psychological Methods}, \textbf{4}(2), 139--157.

\bibitem{Nagin2010} 
Nagin, D. S., \& Odgers, C. L. (2010). Group-based trajectory modeling in clinical research. \textit{Annual Review of Clinical Psychology}, \textbf{6}(1), 109--138.

\bibitem{Olsen2001} 
Olsen, M. K., \& Schafer, J. L. (2001). A two-part random-effects model for semicontinuous longitudinal data. \textit{Journal of the American Statistical Association}, \textbf{96}, 730--745.

\bibitem{Park2008} 
Park, T., \& Casella, G. (2008). The Bayesian lasso. \textit{Journal of the American Statistical Association}, \textbf{103}, 681--686.

\bibitem{PintoCosta2023} 
Pinto da Costa, J. F., Ferreira, F., Mascarello, M., \& Gaio, R. (2023). Clustering of longitudinal trajectories using correlation-based distances. \textit{SN Computer Science}, \textbf{2}(6).

\bibitem{ProustLima2017} 
Proust-Lima, C., Philipps, V., \& Liquet, B. (2017). Estimation of extended mixed models using latent classes and latent processes: The R package LCMM. \textit{Journal of Statistical Software}, \textbf{78}(2), 1--56.

\bibitem{Tibshirani1996} 
Tibshirani, R. (1996). Regression shrinkage and selection via the lasso. \textit{Journal of the Royal Statistical Society: Series B (Methodological)}, \textbf{58}(1), 267--288.

\bibitem{Tobin1958} 
Tobin, J. (1958). Estimation of relationships for limited dependent variables. \textit{Econometrica}, \textbf{26}, 24--36.

\bibitem{Wang2007} 
Wang, H., \& Leng, C. (2007). Unified lasso estimation via least squares approximation. \textit{Journal of the American Statistical Association}, \textbf{102}, 5277--5286.

\bibitem{Yau2001} 
Yau, K. K., \& Lee, A. H. (2001). Zero-inflated Poisson regression with random effects to evaluate an occupational injury prevention programme. \textit{Statistics in Medicine}, \textbf{20}, 2907--2920.

\bibitem{Zhou2023} 
Zhou, J., Zhang, Y., \& Tu, W. (2023). clusterMLD: An efficient hierarchical clustering method for multivariate longitudinal data. \textit{Journal of Computational and Graphical Statistics}, \textbf{32}(3), 1131--1144.

\bibitem{Zou2006} 
Zou, H. (2006). The adaptive lasso and its oracle properties. \textit{Journal of the American Statistical Association}, \textbf{101}(476), 1418--1429.


\end{thebibliography}
\end{document}